\documentclass[conference]{IEEEtran}
\IEEEoverridecommandlockouts
\usepackage{cite}
\usepackage{amsmath,amssymb,amsfonts}
\usepackage{algorithmic}
\usepackage{graphicx}
\usepackage{textcomp}
\usepackage[colorlinks=true, urlcolor=blue]{hyperref}
\usepackage{xcolor}
\usepackage{tikz}
\usepackage{url}
\usepackage{float}
\usepackage{comment}
\usepackage{algorithm}
\usepackage{algorithmic}
\usetikzlibrary{arrows.meta, positioning}
\def\BibTeX{{\rm B\kern-.05em{\sc i\kern-.025em b}\kern-.08em
    T\kern-.1667em\lower.7ex\hbox{E}\kern-.125emX}}
\begin{document}

\title{Construction and Natural Language Querying of a Cybersecurity Knowledge Graph}

\author{

\IEEEauthorblockN{
1\textsuperscript{st} Ines BEN BRAHIM \quad
2\textsuperscript{nd} Mohamed Amine EL MORTAJI \quad
3\textsuperscript{rd} Nada HADDAD
}
\IEEEauthorblockN{
4\textsuperscript{th} Sami REZIG \quad
5\textsuperscript{th} Sofiane TADIMI \quad
6\textsuperscript{th} Mohamed-Lamine MESSAI
}

\IEEEauthorblockA{
\textit{Université Lumière Lyon 2, Université Claude Bernard Lyon 1, ERIC} \\
69007 Lyon, France \\
\textsuperscript{(1)}\texttt{ines.ben-brahim},
\textsuperscript{(2)}\texttt{mohamed-amine.el-mortaji},
\textsuperscript{(3)}\texttt{nada.haddad}, \\
\textsuperscript{(4)}\texttt{sami.rezig},
\textsuperscript{(5)}\texttt{s.tadimi},
\textsuperscript{(6)}\texttt{mohamed-lamine.messai}@univ-lyon2.fr
}

\vspace{0.5cm}

\IEEEauthorblockN{
7\textsuperscript{th} Kamal BENZEKKI
}
\IEEEauthorblockA{
\textit{INSA Lyon, CITI, Inria} \\
69621 Villeurbanne, France \\
\texttt{kamal.benzekki}@insa-lyon.fr, \texttt{kamal.benzekki}@inria.fr
}

}

\maketitle

% ---------------- ABSTRACT ----------------
\begin{abstract}
Cybersecurity vulnerability information is distributed across numerous platforms and databases, making it difficult for researchers and practitioners to obtain a unified and structured understanding of existing threats. This is a critical issue in cybersecurity, where timely access to accurate vulnerability information directly impacts risk assessment and decision-making. While previous work has shown that knowledge graphs are effective for organizing vulnerability data, a major research gap remains in their accessibility, as querying such graphs typically requires expertise in graph query languages like Cypher. This paper aims to address this gap by proposing an approach that combines the construction of a cybersecurity knowledge graph with natural language-based interrogation. The proposed methodology relies on data collected from the National Vulnerability Database (NVD)\cite{nvd} through its REST API and models vulnerabilities, products, vendors, severity metrics, weaknesses, and references using the Labeled Property Graph paradigm in Neo4j. The knowledge graph is deployed on Neo4j Aura Cloud and queried through an AI-assisted interface that translates natural language queries into Cypher language. The key contribution of this work is demonstrating that natural language querying significantly lowers the barrier to interacting with cybersecurity knowledge graphs, enabling more intuitive exploration and analysis of vulnerability data, and thereby enhancing their practical usefulness for a broader range of users in the cybersecurity field.
\end{abstract}
% ---------------- INTRODUCTION ----------------
\section{Introduction}
Cybersecurity has become a critical concern as software systems grow in size, complexity, and interconnectivity. Every year, thousands of new security vulnerabilities are disclosed \cite{nvd}, affecting a wide range of software and hardware products. For researchers and security practitioners, understanding these vulnerabilities is essential for risk assessment, threat analysis, and decision-making. However, vulnerability information is often dispersed across multiple platforms, reports, and databases, making comprehensive analysis time-consuming and error-prone.

Although centralized repositories such as the NVD \cite{nvd} provide standardized vulnerability data, the volume and complexity of this information remain challenging. Vulnerability records include heterogeneous elements such as textual descriptions, severity metrics, affected products, vendors, and references to external advisories. Traditional tabular or document-based representations struggle to capture the rich relationships that exist between these elements, limiting their analytical value \cite{hogan2021knowledge,robinson2015graph}.

Cybersecurity Knowledge Graphs (CKG) \cite{zhao2023survey}, \cite{benzekki2026empowering} offer a promising solution by representing cybersecurity data as interconnected entities and relationships, enabling a more explicit and semantically rich view of the vulnerability landscape. By modeling vulnerabilities, products, vendors, and weaknesses as nodes linked through meaningful relationships, knowledge graphs support advanced exploration and reasoning.
 This expressiveness, however, creates a barrier to adoption. Because querying a knowledge graph typically requires expertise in languages like Cypher, which limits its utility for broader cybersecurity analysis.

Recent advances in AI, particularly large language models, have opened new possibilities for lowering this barrier. Natural language interfaces for graph databases allow users to query complex data structures without explicit knowledge of graph query syntax. Neo4j Aura, a managed cloud platform for graph databases, integrates an AI-assisted search mechanism that translates natural language queries into Cypher language \cite{bratanic2024graph}, making graph-based analysis more accessible.

The core contribution of this paper is a complete, end-to-end demonstration of a natural language-queryable cybersecurity knowledge graph. We describe the data collection process, graph modeling choices, and cloud deployment, and we evaluate the use of AI-assisted natural language querying for exploring vulnerability information. The results highlight the benefits of combining knowledge graphs with natural language interfaces for exploratory cybersecurity analysis, while also identifying current limitations when addressing complex and multi-hop queries.

The remainder of this paper is structured as follows. Section 2 reviews the related works and section 3 details our methodology for constructing and deploying the knowledge graph. Then, section 4 presents our evaluation results. Next, section 5 discusses the findings and limitations. Finally, section 6 concludes and suggests future research directions.

%New Related wok section 
% ---------------- STATE OF THE ART ----------------
\subsection{Related Work}

The growing volume and complexity of cybersecurity data have led several researchers to adopt knowledge graphs as a way to structure vulnerability information. These graphs enable richer semantic relationships and more flexible querying than traditional databases. However, most existing solutions still require users to write queries in languages like Cypher, which limits their accessibility. This section reviews representative approaches to cybersecurity knowledge graph construction and compares them with our work, focusing on whether they support natural language interfaces.

A notable effort in this direction comes from Falcarin and Dainese \cite{falcarin2024cybergraph}, who propose CyberGraph, a tool that automatically builds a cybersecurity knowledge graph by integrating public repositories such as CVE, CWE, CAPEC, CPE, and CVSS. The graph is stored in Neo4j and follows a manually designed schema that preserves the original data structures. A clear strength of CyberGraph is its scale. It contains over 735,000 nodes and 1.3 million relationships, besides it supports rich cross‑domain queries across vulnerabilities, weaknesses, and attack patterns. The authors also provide utility scripts for incremental updates. On the other hand, the system requires users to write Cypher queries explicitly; there is no natural language interface. This means that security analysts without graph query training cannot easily explore the data. Moreover, the manual ontology design, while precise, may become a bottleneck when the schema needs to evolve rapidly.

A different strategy is taken by Host et al. \cite{host2023kg}, who present an automatic method to build a vulnerability knowledge graph directly from textual descriptions in the NVD. Their pipeline uses SecBERT and Averaged Perceptron for named entity recognition, rule‑based relation extraction, and TuckER embeddings for entity prediction (to recover missing software or weakness types). One advantage of this work is that it reduces manual effort by extracting entities and relations directly from unstructured text. The use of knowledge graph embeddings for completion is also a valuable contribution, and the evaluation covers about 175,000 CVEs. However, the fully automatic extraction introduces noise where the precision for relation extraction is only 0.77. More importantly, the authors do not address natural language querying at all. So, users must work directly with the graph structure. The relation extraction is based on a fixed ontology and simple word‑order rules, which may miss more complex or implicit relationships.

While Host et al. focus on automation, Li et al. \cite{cyberkg2023} place greater emphasis on data quality and graph assessment. They develop a framework for constructing cybersecurity knowledge graphs and assessing their quality, manually creating the CS13K dataset (13,027 triples, 4,494 entities, 12 relations) and extending the UCO ontology to 16 classes. They also propose AttTucker, a Transformer‑based model that evaluates triple confidence, and show that incorporating path‑level information improves quality assessment. A major strength is the high quality of the manually annotated dataset, which serves as a reliable benchmark. The quality assessment model is novel and effective, achieving 0.947 accuracy on noisy data. Nevertheless, the work does not provide any natural language querying capability. Even with high‑quality triples, users still need Cypher expertise to interact with the graph. Furthermore, the ontology and graph construction are manual, which ensures correctness but may not scale to fast‑changing threat landscapes without significant human effort.

An earlier and more foundational contribution comes from Jia et al. \cite{practicalkg2018}, who propose a practical approach based on a quintuple model (concept, instance, relation, properties, rule). They use machine learning (Stanford NER) to extract entities from structured and unstructured data and apply path‑ranking algorithms for knowledge deduction, inferring new attributes and relationships. The main contribution is a clear framework for ontology construction and rule‑based reasoning. The use of CRF‑based NER is well‑justified, and the deduction rules add some reasoning capability. On the downside, the graph construction is semi‑automatic and still heavily relies on manual ontology engineering. The NER performance (F1 around 0.83 for the best model) leaves room for improvement. Crucially, there is no natural language interface, thus, querying the graph requires knowledge of the schema and Cypher. This work predates the widespread availability of large language models and does not leverage modern NLU techniques.

As the above review shows, all previous approaches focus primarily on knowledge graph construction—whether manual, automatic, or semi‑automatic and none integrates a natural language querying interface. Table~\ref{tab:construction_comparison} summarises this gap. Cybersecurity practitioners, who often lack Cypher skills, are therefore unable to fully exploit these graphs.

\begin{table}[htbp]
\centering
\caption{Comparison of approaches on cybersecurity knowledge graph construction and natural language querying.}
\label{tab:construction_comparison}
\resizebox{\columnwidth}{!}{
\begin{tabular}{lcc}
\hline
\textbf{Article} & \textbf{Graph construction} & \textbf{LLM for NL querying} \\
\hline
Falcarin and Dainese (CyberGraph, 2024) \cite{falcarin2024cybergraph} & Manual & No \\
Host et al. (2023) \cite{host2023kg} & Automatic & No \\
Li et al. (2024) \cite{cyberkg2023} & Manual & No \\
Jia et al. (2018) \cite{practicalkg2018} & Semi-automatic & No \\
\hline
\textbf{Our work} & \textbf{Manual} & \textbf{Yes} \\
\hline
\end{tabular}
}
\end{table}
In contrast, our work directly addresses this gap by combining a cybersecurity knowledge graph (built from NVD data and modeled in Neo4j) with an AI‑assisted natural language interface. By leveraging Neo4j Aura’s built‑in LLM‑based translation, users can express their information needs in plain English without writing any Cypher. Our evaluation shows that natural language queries successfully handle simple to moderately complex tasks such as one‑hop traversals, filtering, and basic aggregations, thus lowering the barrier to exploratory vulnerability analysis. While we acknowledge current limitations for multi‑hop queries and highly compositional patterns, our approach demonstrates the first end‑to‑end, natural language queryable cybersecurity knowledge graph that is both practical and accessible to a broad range of security researchers and professionals.

\section{Methodology}

This methodology is divided into two main phases. The first phase focuses on the construction of the cybersecurity knowledge graph, including data collection and graph modeling. The second phase addresses the interrogation of the graph using natural language queries through Neo4j Aura.

% =========================================================
% 1. CONSTRUCTION OF THE KNOWLEDGE GRAPH
% =========================================================
\subsection{Construction of the Knowledge Graph}

The construction of a cybersecurity knowledge graph involves transforming structured vulnerability data into interconnected semantic entities. This makes it easier to analyze and make decisions by providing an explicit and exploitable representation of the connections between security vulnerabilities, affected products, and their characteristics.

\subsubsection*{Data Collection}

A reliable and comprehensive dataset was required to construct the graph. We chose the NVD \cite{nvd} which offers standardized information on security vulnerabilities identified using the CVE \cite{cve} standard, which is widely recognized in cybersecurity research and practice.

Data were retrieved via the public NVD REST API (version 2.0) in JSON format. For this project, an initial dataset of 2,000 CVE records was saved locally. Each record in the dataset includes a unique \textit{CVE identifier}, a textual \textit{description} of the vulnerability, the \textit{affected products} (software or hardware impacted by the vulnerability), \textit{severity metrics} (including CVSS scores), \textit{weaknesses} (the underlying root causes categorized by the Common Weakness Enumeration (\textit{CWE}) taxonomy, such as buffer overflow, injection, or improper input validation), \textit{references} to external advisories, patches, or documentation, and \textit{metadata} such as \texttt{resultsPerPage}, \texttt{startIndex}, \texttt{totalResults}, and the API version.

The collection process was automated by building a pipeline, that extracted, formatted, and stored the data for later ingestion into the knowledge graph. This pipeline guarantees reproducibility and allows simple updates when new CVEs are published.

Since CVE records can contain multiple versions of CVSS metrics, the ingestion process prioritizes CVSS v3.1 if available, followed by CVSS v3.0 and CVSS v2 as fallback options. This hierarchical selection strategy ensures the use of the most recent and expressive severity metrics. Furthermore, separation between data acquisition and graph ingestion enables the system to scale naturally. New CVE batches may be retrieved and integrated without the need to rebuild the entire graph, making this approach suitable for continuous vulnerability monitoring scenarios.

\subsubsection*{Knowledge Graph Design}

Important cybersecurity entities were modeled as nodes in the knowledge graph, with their semantic relationships represented as edges.  The design follows the Labeled Property Graph (LPG) paradigm \cite{neo4j}, which is supported by graph database systems like Neo4j. Each node type contains specific properties, while edges  specify how nodes relate to one another.

\paragraph*{Global Knowledge Graph Overview} 

To provide a comprehensive understanding of the overall structure, Figure 1 presents a global view of the cybersecurity knowledge graph. The graph illustrates the main entities involved in vulnerability management, including CVEs, affected products, vendors, severity metrics, weakness categories (CWE), and external references, along with the relationships that connect them.

This global representation highlights how vulnerabilities are linked to impacted products and vendors, how their severity is characterized through standardized metrics, and how they are associated with underlying weakness categories and external documentation. The holistic view serves as a foundation for efficient knowledge exploration and pattern-based querying using Neo4j.

\begin{figure}[h]
    \centering
    \includegraphics[width=\columnwidth]{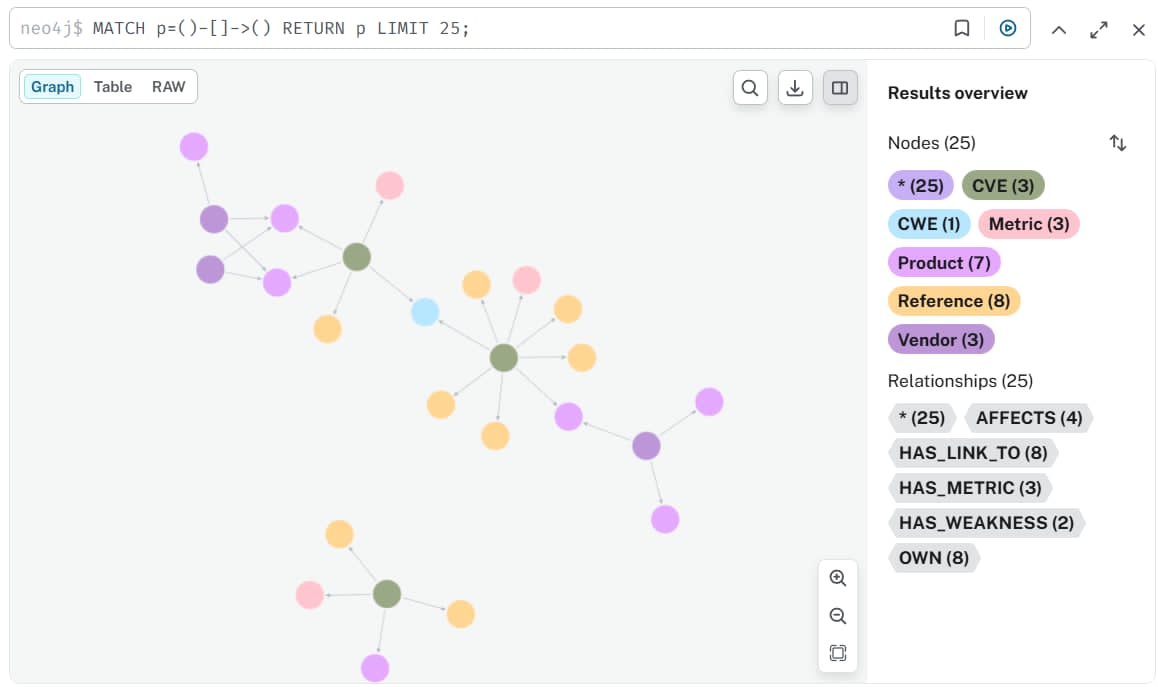} % fills one column
    \caption{Global structure of the cybersecurity knowledge graph}
    \label{fig:myfigure}
\end{figure}

\paragraph{Graph Database and Storage Technology (Neo4j)}
The constructed knowledge graph is stored and managed using Neo4j, a native graph database management system that supports the Labeled Property Graph (LPG) model, in which nodes and relationships can carry labels and properties \cite{neo4jdocs}.

Neo4j is suited for cybersecurity applications since it efficiently represents complex relationships between vulnerabilities, affected products, vendors, and weakness categories. Its declarative query language, Cypher, handles expressive pattern-based queries, facilitating vulnerability analysis and scalable integration of recently released CVE records from the National Vulnerability Database (NVD).

\paragraph{Nodes (Entities)}

The primary entities represented as nodes in the graph include CVE, Metric, Product, Vendor, CWE, and Reference. A \textit{CVE} node represents a unique security vulnerability, with properties including \texttt{id}, \texttt{description}, \texttt{publishedDate}, and \texttt{lastModifiedDate}. A \textit{Metric} node contains severity metrics based on the CVSS scoring system, storing properties such as \texttt{baseScore} and \texttt{vector}. \textit{Product} nodes denote the software or hardware impacted by a given CVE, with a property for the product name, while \textit{Vendor} nodes represent the entity responsible for a product, storing the vendor name. \textit{CWE} nodes model vulnerability root causes using the Common Weakness Enumeration (CWE) taxonomy; each CVE is linked to one or more CWE nodes representing its underlying weakness type. This allows vulnerabilities to be grouped and analyzed based on common design or implementation flaws rather than isolated identifiers. Finally, \textit{Reference} nodes capture external references associated with CVEs, such as vendor advisories, security bulletins, and documentation links, modeling them as first-class entities. This design choice enables multiple vulnerabilities to reference the same external source and supports traceability between vulnerabilities and mitigation guidance.

\paragraph{Relationships (Edges)}

%\begin{itemize}
%    \item \texttt{(CVE) -[HAS\_METRIC]-> (Metric)}
%\end{itemize}

% --- CVE ↔ Metric ---
\begin{figure}[h!]
    \centering
    \begin{tikzpicture}[
        node distance=1.2cm, auto,
        box/.style={draw, rectangle, rounded corners, minimum width=2.0cm, minimum height=0.7cm, align=center, font=\scriptsize, fill=blue!10},
        relationship/.style={->, thick, >=Stealth}
    ]
        \node[box] (CVE) {CVE\\(Vulnerability)};
        \node[box, right=2.5cm of CVE] (Metric) {Metric\\(Severity Scores)};
        \draw[relationship] (CVE) -- node[above, font=\scriptsize] {HAS\_METRIC} (Metric);
    \end{tikzpicture}
    \caption{CVE connected to its severity metrics.}
\end{figure}
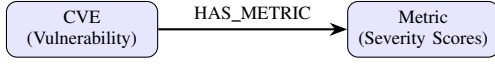

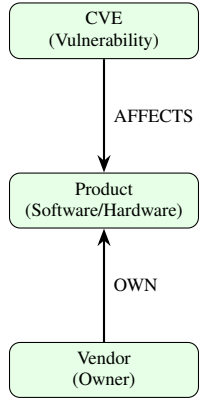
\begin{figure}[H]
    \centering
    \begin{tikzpicture}[
        node distance=1.2cm, auto,
        box/.style={draw, rectangle, rounded corners, minimum width=2.5cm, minimum height=0.7cm, align=center, font=\scriptsize, fill=green!10},
        relationship/.style={->, thick, >=Stealth}
    ]
        \node[box] (CVE) {CVE\\(Vulnerability)};
        \node[box, below=1.5cm of CVE] (Product) {Product\\(Software/Hardware)};
        \node[box, below=1.5cm of Product] (Vendor) {Vendor\\(Owner)};
        \draw[relationship] (CVE) -- node[right, font=\scriptsize] {AFFECTS} (Product);
        \draw[relationship] (Vendor) -- node[right, font=\scriptsize] {OWN} (Product);
    \end{tikzpicture}
    \caption{CVE affecting products and their corresponding vendors.}
\end{figure}

%\begin{itemize}
%    \item \texttt{(CVE) -[HAS\_WEAKNESS]-> (CWE)}
%\end{itemize}

% --- CVE ↔ CWE ---
\begin{figure}[h!]
    \centering
    \begin{tikzpicture}[
        node distance=1.5cm, auto,
        box/.style={draw, rectangle, rounded corners, minimum width=2.0cm, minimum height=0.7cm, align=center, font=\scriptsize, fill=orange!10},
        relationship/.style={->, thick, >=Stealth}
    ]
        \node[box] (CVE) {CVE\\(Vulnerability)};
        \node[box, right=2.5cm of CVE] (CWE) {CWE\\(Weakness Category)};
        \draw[relationship] (CVE) -- node[above, font=\scriptsize] {HAS\_WEAKNESS} (CWE);
    \end{tikzpicture}
    \caption{CVE associated with its weakness category (CWE).}
\end{figure}
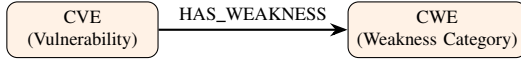

%\begin{itemize}
%    \item \texttt{(CVE) -[HAS\_LINK\_TO]-> (Reference)}
%\end{itemize}

% --- CVE ↔ Reference ---
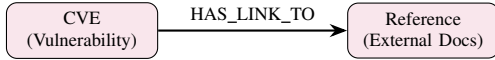
\begin{figure}[h!]
    \centering
    \begin{tikzpicture}[
        node distance=1.5cm, auto,
        box/.style={draw, rectangle, rounded corners, minimum width=2.0cm, minimum height=0.7cm, align=center, font=\scriptsize, fill=purple!10},
        relationship/.style={->, thick, >=Stealth}
    ]
        \node[box] (CVE) {CVE\\(Vulnerability)};
        \node[box, right=2.5cm of CVE] (Reference) {Reference\\(External Docs)};
        \draw[relationship] (CVE) -- node[above, font=\scriptsize] {HAS\_LINK\_TO} (Reference);
    \end{tikzpicture}
    \caption{CVE connected to external documentation or advisories.}
\end{figure}

% =========================================================
% 2. GRAPH INTERROGATION
% =========================================================
\subsection{Graph Interrogation Using Natural Language}

The increasing adoption of knowledge graphs has highlighted the need for more accessible querying tools. Neo4j Aura, the managed cloud offering of Neo4j, introduces an AI-assisted search mechanism to translate natural language into Cypher queries, bridging the gap for users not proficient in graph-specific syntax. This paper examines this feature, focusing on its conceptual mechanism, capabilities, and inherent limitations, particularly when dealing with complex or dense knowledge graphs. The feature is well suited for exploratory analysis and educational use but reveals limitations when faced with complex, multi-hop relationship reasoning.

\subsubsection*{Conceptual Mechanism of AI Search}

\begin{figure}[h!]
\centering
\begin{tikzpicture}[
    node distance=0.9cm,
    >=Stealth,
    box/.style={
        draw,
        rectangle,
        rounded corners,
        minimum width=2.6cm,
        minimum height=0.8cm,
        align=center,
        font=\footnotesize
    }
]

% Main vertical flow
\node (NL) [box] {NL Query\\(Input)};
\node (LLM) [box, below=of NL] {LLM / Semantic\\Parser};
\node (Cypher) [box, below=of LLM] {Cypher Query\\(Generated)};
\node (Neo4j) [box, below=of Cypher] {Neo4j Engine\\(Execution)};
\node (Result) [box, below=of Neo4j] {Graph Result\\(Nodes / Relations)};

% Side metadata box
\node (Schema) [box, right=1.2cm of LLM, fill=gray!10] 
{Graph Schema\\Metadata};

% Arrows
\draw[->, thick] (NL) -- (LLM);
\draw[->, thick] (LLM) -- (Cypher);
\draw[->, thick] (Cypher) -- (Neo4j);
\draw[->, thick] (Neo4j) -- (Result);

\draw[->, thick, dashed] (Schema.west) -- (LLM.east);

\end{tikzpicture}
\caption{Conceptual Pipeline of Neo4j Aura's AI-Assisted Search}
\end{figure}
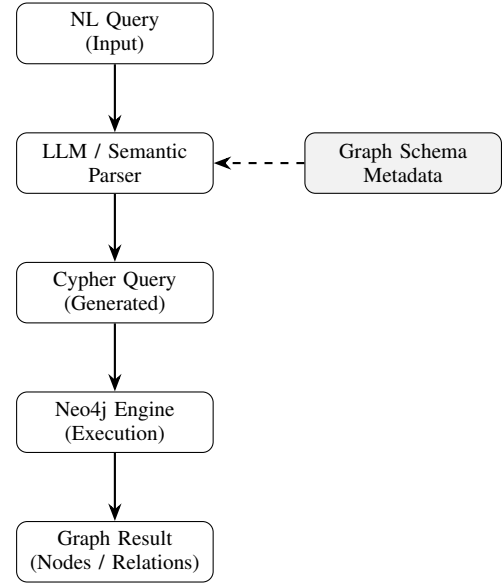

The Large Language Model (LLM) interprets the Natural Language (NL) input, using the Graph Schema Metadata to constrain the output and generate a valid Cypher query (Semantic Parsing). Before discussing natural language querying, it is useful to briefly outline how Neo4j Aura enables AI-assisted search at a conceptual level . Although the internal implementation details are not publicly disclosed, the AI search feature can be understood as a pipeline that combines schema awareness with large language model–based query generation .

The system has access to metadata about node labels, relationship types, and properties defined in the graph schema . When a user provides a natural language query, this input is interpreted in the context of the schema and mapped to an equivalent Cypher query that aims to retrieve the requested information . This process resembles semantic parsing, where unstructured text is transformed into a structured query language, with constraints imposed by the graph model to reduce ambiguity and invalid queries .

\subsubsection*{Cloud Deployment of an Existing Knowledge Graph}

Once the knowledge graph has been developed and validated locally, deploying it to Neo4j Aura Cloud requires only minimal modifications. The data model, constraints, and Cypher scripts remain unchanged. The main difference lies in the connection configuration. Instead of connecting to a local Neo4j instance, the application or script must be configured to use the URI, username, and password associated with the Neo4j Aura instance.

To obtain these credentials, a new Aura database instance must first be created through the Neo4j Aura console. After creation, selecting the \emph{Connect} option and navigating to the \emph{Developer Hub}, then choosing Python, provides the exact connection details required. By replacing the local connection parameters with these cloud credentials and executing the same ingestion script described in the previous document, the database is populated in the Aura environment without further changes.

\subsubsection*{Natural Language Interface for Graph Exploration}

\begin{figure}[h]
    \centering
    \includegraphics[width=\columnwidth]{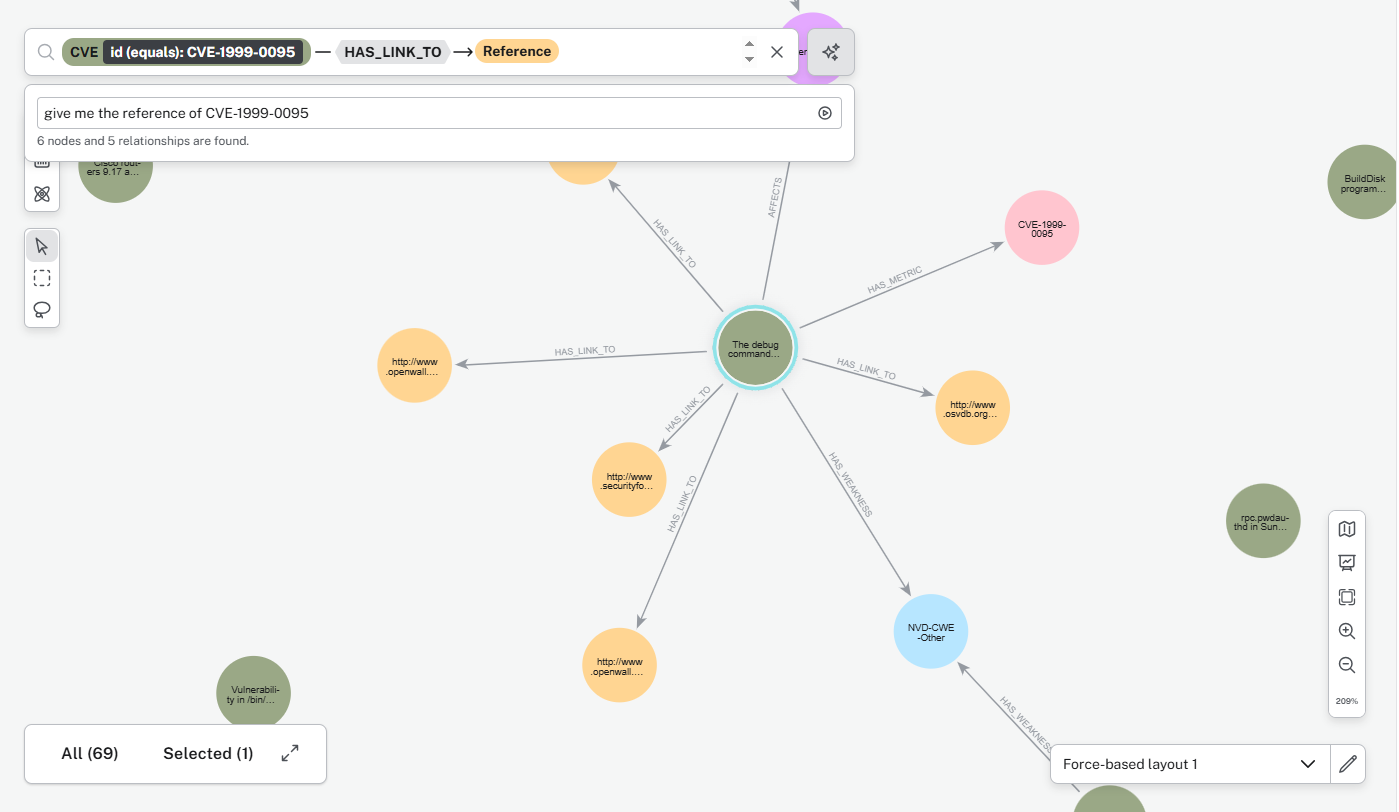} % fills one column
    \caption{Query through Neo4j built in natural language processor}
    \label{fig:myfigure}
\end{figure}

After the database has been successfully deployed, Neo4j Aura provides an interface for querying the graph using natural language. This functionality is accessible through the \emph{Explore} tab of the Aura console.
Within the search input field, users can enable the AI search option, which allows them to express queries in plain English rather than Cypher. The user simply describes the information they wish to retrieve, and the system automatically generates the corresponding Cypher query. For example, a query requesting all entities related to a given node type or a specific relationship pattern can be expressed in natural language, and the generated Cypher can be inspected, refined, or executed directly.

\subsubsection*{Limitations of AI-Assisted Natural Language Queries}

Despite its convenience, the AI-assisted natural language querying feature has notable limitations. In practice, it performs best for relatively simple retrieval tasks involving a small number of node types and relationships. The quality of the generated Cypher may degrade as queries grow more complex, particularly when multiple hops, aggregations, or intricate relationship constraints are involved. The system may produce incomplete queries, omit necessary constraints, or misinterpret the intended traversal
1 of the graph. This limitation is particularly evident in dense knowledge graphs where many node labels and relationships coexist, increasing semantic ambiguity in natural language descriptions.

\subsection{Code and Reproducibility Resources}

All source code, data preprocessing scripts, Cypher queries, and the final version of this article are publicly available in the project repository:

{\small\noindent\url{https://github.com/NadaHaddad/graph_security}}

This repository provides full reproducibility for data ingestion, graph construction, AI versus expert query evaluation, and all experiments presented in this work.

% ---------------- RESULTS ----------------
\section{Results}
The resulting knowledge graph successfully represents vulnerability data as interconnected entities. Neo4j enables efficient traversal of relationships, allowing complex queries that would be difficult to express using relational databases. Natural language querying further improves usability by allowing users to retrieve information without writing Cypher queries manually.  Table~\ref{tab:ai_cypher_evaluation} summarizes the results.

\begin{table}[H]
\centering
\caption{Evaluation of AI-generated Cypher queries using natural language prompts.}
\label{tab:ai_cypher_evaluation}
\resizebox{\columnwidth}{!}{%
\begin{tabular}{|c|p{8.2cm}|c|}
\hline
\textbf{ID} & \textbf{Natural Language Prompt} & \textbf{Result} \\
\hline
1 & Show the top 10 vendors with the most CVEs. & Partial \\
\hline
2 & List the CVEs with a base score of 9.0 or higher and show the affected products. & Partial \\
\hline
3 & Find CVEs that affect products from more than one vendor. & Success \\
\hline
4 & Show the number of CVEs published each year. & Partial \\
\hline
5 & List the top 5 most frequent CWE weakness categories in the database. & Partial \\
\hline
6 & Find all CVEs that have a reference link containing the word ``github''. & Success \\
\hline
7 & List CVEs that do not have an associated severity metric. & Success \\
\hline
8 & List all products affected by the CVE ``CVE-1999-0002''. & Success \\
\hline
9 & Show all CVEs published in 1999 that mention the word ``buffer'' in the description. & Success \\
\hline
10 & For each vendor, show how many products they own and how many CVEs affect those products. & Partial \\
\hline
\end{tabular}%
}
\end{table}
\smallskip
\textit{Note: “Partial” indicates that the generated Cypher query was syntactically valid but required minor corrections to achieve full semantic correctness.}

\subsection{Evaluation of AI-Generated Cypher Queries}

To assess the reliability of Neo4j Aura’s AI-assisted natural language querying, 
we designed an experiment consisting of ten representative prompts covering 
simple retrieval tasks, multi-hop traversals, aggregations, and filtering operations. 
For each prompt, the Cypher query generated by the AI was compared against a 
hand-crafted expert query and evaluated based on correctness, completeness, 
and schema alignment.

Overall, the system performed remarkably well on direct one-hop traversals 
(e.g., retrieving affected products or listing CVEs with specific attributes), 
consistently producing correct Cypher queries without requiring manual refinement. 
Prompts involving aggregation or filtering also yielded accurate outputs, 
demonstrating that the model handles common analytical patterns such as 
\texttt{COUNT}, \texttt{ORDER BY}, and substring extraction.

Performance decreased slightly for structurally complex prompts requiring 
multi-hop reasoning or multi-level grouping, where the AI occasionally omitted 
distinguishing constraints (e.g., unique vendor counts). Nevertheless, even in 
these cases, the generated queries remained syntactically valid and required 
only minimal adjustments. These observations confirm that natural language 
interfaces significantly improve the accessibility of knowledge graph querying, 
while highlighting their current limitations for deeply compositional graph patterns. These observations provide a concrete foundation for the broader implications and limitations discussed in the following section.

% ---------------- DISCUSSION ----------------
\section{Discussion}
The proposed approach demonstrates the advantages of combining knowledge graphs with natural language interfaces for cybersecurity analysis, improving accessibility for users who are not yet proficient in Cypher. In practice, the natural language querying capabilities provided by Neo4j Aura are well suited for exploratory analysis, basic retrieval tasks, and educational use cases. However, the accuracy and reliability of the generated queries depend strongly on the quality of the underlying graph schema and the language model’s understanding of domain-specific terminology. As query complexity increases, particularly in the presence of multi-hop relationships or dense graph structures, manual Cypher queries or programmatic approaches remain more reliable and may require user refinement. A promising direction for future work would involve integrating external natural language processing pipelines, such as Python-based scripts interacting with large language model APIs, to enable finer control over prompt engineering, schema grounding, and query validation, potentially mitigating some of the limitations observed in the built-in AI-assisted search.

% ---------------- CONCLUSION ----------------
\section{Conclusion}

This paper presented CyberGraph, a cybersecurity knowledge graph constructed from publicly available vulnerability data and queried through natural language. By combining the graph-based capabilities of Neo4j with the use of large language models to make a Cypher query from natural language, the proposed approach facilitates vulnerability exploration and provides a more intuitive means of analyzing complex cybersecurity information. The results highlight the potential of this integration to improve access to structured vulnerability knowledge and support analytical tasks in a more user-friendly manner. 

Future work will investigate the enrichment of CyberGraph with heterogeneous cybersecurity sources, including threat intelligence feeds, attack pattern repositories, and remediation knowledge, in order to improve both coverage and contextualization. Another important direction concerns the evaluation and enhancement of natural language query translation, particularly through the design of domain-specific prompting strategies, schema-aware reasoning, and benchmark-based performance assessment.

% ---------------- REFERENCES ----------------
\bibliographystyle{plain}  
\bibliography{references}

\end{document}